\documentclass[]{aastex631}

\submitjournal{RNAAS}

\shorttitle{Grid of upper atmosphere models revised}
\shortauthors{Kubyshkina et al.}

\begin{document}

\title{Extending a grid of hydrodynamic planetary upper atmosphere models}

\author[0000-0001-9137-9818]{Daria I. Kubyshkina}
\affiliation{Trinity College Dublin \\
Dubline-2 College Green \\
Dublin, Ireland}

\author{Luca Fossati}
\affiliation{Space Research Institute\\
8042 Graz, Austria}



\begin{abstract}
In this research note, we outline the extension of the grid of upper atmosphere models first presented in \citet{kubyshkina2018grid}. The original grid is based on a 1D hydrodynamic model and consists of about 7000 models covering planets of the size from Earth to twice Neptune at orbits corresponding to equilibrium temperatures between 300 and 2000 K around solar-like (0.4 to 1.3 solar mass) stars. The extended and revised grid of models consists of 10235 points and covers a planetary mass range of up to 109 Earth masses, which allows one to outline the transition between low- and high-gravity hot planets in short orbital separations. We prepared the interpolation tool allowing one to use the grid to define the mass-loss of a planet that falls into the parameter range of the grid. We provide a comparison of our results to common analytical models.

\end{abstract}

\keywords{Hydrodynamics (1963) --- Exoplanet atmospheres (487) --- Upper atmosphere (1748) --- Exoplanet atmospheric variability (2020)}


\section{Grid of upper atmosphere models} \label{sec:intro}

In \citet{kubyshkina2018grid} we presented a grid of 1D hydrodynamic models of planetary upper atmospheres based on the upgraded model of \citet{erkaev2016MNRAS.460.1300E}. The model considers pure hydrogen atmosphere and accounts for dissociation, recombination and ionization, and cooling processes such as $Ly\alpha-$ and $H_3^+-$cooling. The atmosphere is heated by stellar high-energy (X-ray + EUV, XUV) flux, where the integrated X-ray and EUV fluxes are assumed to be emitted at the single wavelengths, of 5 and 60 nm, respectively. The heating efficiency was assumed to be 15\% throughout the grid. 

The original grid covered the following ranges of planetary and stellar parameters. For planets, we considered masses ($M_{pl}$) of 1 to 39 $M_{\oplus}$ and radii ($R_{pl}$) of 1 to 10 $R_{\oplus}$. As host stars, we considered those in the mass range 0.4-1.3 $M_{\odot}$ presenting a wide spread in XUV luminosities, scaled in accordance to the stellar mass. We considered orbits corresponding to equilibrium temperatures ($T_{eq}$) between 300 and 2000 K, for which the specific orbital distances were calculated on the basis of stellar evolution models by \citet{yi2001stellar-models}. We excluded from consideration planets with densities below 0.03 $g/cm^3$ and reduced Jeans escape parameter \citep{fossati2017aeronomical_constr}

\begin{equation}\label{eq:lambda}
\Lambda = \frac{G M_{\rm pl}m_{\rm H}}{k_{\rm b}T_{\rm eq}R_{\rm
pl}}
\end{equation}

\noindent higher than 80. 

Further details on the hydrodynamic model and the grid structure can be found in \citet{kubyshkina2018grid}.

\section{Grid extension} \label{sec:extension}

In the present version of the grid, we have extended the planetary mass range up to $\sim$109 Earth masses and included $\Lambda$ values up to $\sim$150, which allows one representing better the spread of escape rates, particularly at close-in orbits. The upgraded tables for escape rates and the Python routine allowing to interpolate within the grid are available in the Zenodo repository at https://doi.org/10.5281/zenodo.4643823. 

In top panel of Figure~1, we present the mass-loss rates obtained for planets present in the grid (black) against $\Lambda$ and compare them to the predictions of frequently used (semi) analytical approaches. First, we consider the most common energy-limited approximation \citep{watson1981energy-limited} that considers atmospheric escape fully induced by the stellar XUV radiation, absorbed at a certain effective radius $R_{eff} > R_{pl}$. For further simplification, it is often assumed that $R_{eff}$ is equal to the planetary radius. We consider this case (green points), and the energy-limited estimate obtained by adopting $R_{eff}$ from our hydrodynamic simulations (blue points). Second, we compare our results to the predictions of the core-powered mass-loss model (magenta points), which considers the atmospheric escape induced by the cooling luminosity of a planetary core. For the latter, we follow the scheme described in \citet{gupta_schlichting2020core-powered} employing the same atmospheric parameters at the photosphere radius of a planet as for our hydrodynamic models.

\begin{figure}[h!]
\plotone{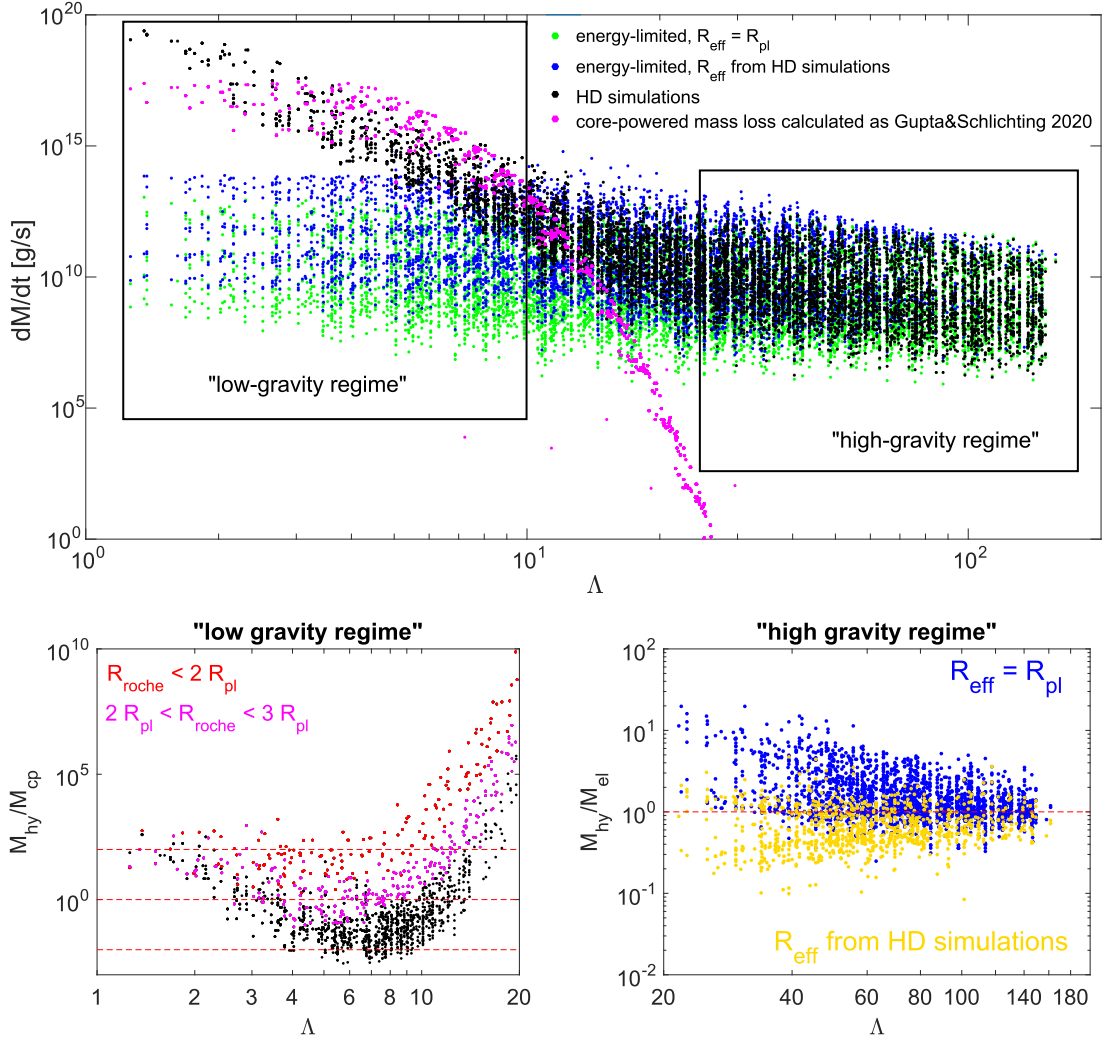}\label{fig::1}
\caption{Top panel: The atmospheric mass-loss rates of the grid planets as given by hydrodynamic models (black); by the energy-limited approximation assuming $R_{eff} = R_{pl}$ (green) or adopting $R_{eff}$ from the hydrodynamic models (blue); and by core-powered mass-loss model assuming the lower boundary conditions as set in the grid (magenta).
Bottom left: The relation of the mass-loss rates given by the hydrodynamic models to the core-powered mass-loss for low-gravity planets. Color code indicates the size of the Roche lobe as shown in the plot. Red dashed lines show the ratios of 0.01, 1, and 100.
Bottom right: The relation of the escape rates given by the hydrodynamic models to the energy-limited mass-loss assuming $R_{eff} = R_{pl}$ (blue) or $R_{eff}$ predicted by hydrodynamic models (yellow), for high gravity planets. Red dashed line shows $\dot{M}_{hy}/\dot{M}_{el} = 1$.
All values are shown against the reduced Jeans escape parameter (Eq. \ref{eq:lambda}).}
\end{figure}

Figure~1 indicates the presence of two mass-loss regimes, as discussed in detail in \citet{kubyshkina2018approx}. The low-mass (and low $\Lambda$) planets experience very high atmospheric escape, which though damps fast with increasing planetary mass. The escape in this region of the parameter space is mainly caused by low gravity and high temperature of planets and depends weakly on stellar XUV. We refer to it as the ``low-gravity regime''. For more massive planets, instead, the mass-loss rates are lower and depend strongly on the amount of XUV received from the star (nearly linear at the high-mass end of the distribution), and change slower with increasing $M_{pl}$. We refer to this as the ``high-gravity regime''.

At a first glance, it looks like the mass-loss rates given by hydrodynamic simulations are consistent with those given by the energy-limited approximation in the ``high-gravity regime'', and are comparable to the core-powered mass-loss in the ``low gravity regime''. Closer inspection, however, reveals significant differences. In the bottom panels of Figure~1, we show the ratio of the escape rates predicted by hydrodynamic modeling to those predicted by core-powered mass-loss for planets in the ``low gravity regime'' (left), and to those predicted by the energy-limited formula for planets in the ``high gravity regime'' (right). 

In the ``low gravity regime'', most of the $\dot{M}_{hy}/\dot{M}_{cp}$ values lie within 0.01-100 for planets with $\Lambda\lesssim15$, and the ratio increases steeply up to a few orders of magnitude at larger $\Lambda$, as planets get into the transition region towards the ``high gravity regime'' and the stellar irradiation starts to become more relevant. Throughout, the ratio decreases with increasing planetary Roche lobe. Thus, for most planets with the Roche radius ($R_{roche}$) larger than 3~$R_{pl}$ (black points) and $\Lambda\leq 10$, the escape predicted by hydrodynamic models is smaller than the core-powered mass-loss, and both are below Bondi-limited mass-loss, which is considered as a physical limit of escape in \citet{gupta_schlichting2020core-powered}. This looks natural, as the core-powered mass-loss in the considered formulation assumes that all the cooling luminosity of a planet is spent on driving escape, providing therefore the upper limit of escape. With decreasing $R_{roche}$, the stellar tides become important. As a proxy, it is convenient to consider the coefficient $K$ introduced in \citet{erkaev2007roche-lobe} to account for Roche lobe effects when using the energy-limited approximation: $K=\frac{(\eta-1)^2(2\eta+1)}{2\eta^3} < 1$, where $\eta = \frac{R_{roche}}{R_{pl}}$, is inversely proportional to the mass-loss rate. It is easy to see, that at $R_{roche} = 3$ (border between black and magenta points), $K \sim 0.5$, and for $R_{roche} \leq 2$ (red points) $K$ changes between $\sim 10^{-6}$ and $\sim 0.3$, for most of the points being above 0.01. This range of values is very similar to what is presented in Figure~1 for $\dot{M}_{hy}/\dot{M}_{cp}$ at $R_{roche} \leq 2$ and $\Lambda\lesssim 10$. 

In the ``high gravity regime'', most of the escape is driven by the stellar XUV. When comparing to the energy-limited escape, however, $\dot{M}_{hy}/\dot{M}_{el}$ presents a wide spread up to about an order of magnitude. When considering the widely applied simplification of $R_{eff} = R_{pl}$ (blue points in bottom right panel of Figure~1), energy-limited escape in general significantly underestimates the escape. However, including the realistic effective radii of the XUV absorption from hydrodynamic simulations does not solve the problem, and leads instead to the overestimation of escape, as the energy-limited escape assumes that all the absorbed XUV flux is spent on driving the escape, while in reality part of it is spent on driving the atmospheric chemistry. The applicability limits of energy-limited approximation were considered in detail by Krenn et al., 2021 (under revision). 

\begin{acknowledgments}
This work was supported by Austrian Forschungsfoerderungsgesellschaft FFG project “TAPAS4CHEOPS” P853993.
\end{acknowledgments}

%





\bibliography{grid_update.bib}{}

\begin{thebibliography}{}
\expandafter\ifx\csname natexlab\endcsname\relax\def\natexlab#1{#1}\fi
\providecommand{\url}[1]{\href{#1}{#1}}
\providecommand{\dodoi}[1]{doi:~\href{http://doi.org/#1}{\nolinkurl{#1}}}
\providecommand{\doeprint}[1]{\href{http://ascl.net/#1}{\nolinkurl{http://ascl.net/#1}}}
\providecommand{\doarXiv}[1]{\href{https://arxiv.org/abs/#1}{\nolinkurl{https://arxiv.org/abs/#1}}}

\bibitem[{{Erkaev} {et~al.}(2007){Erkaev}, {Kulikov}, {Lammer}, {Selsis},
  {Langmayr}, {Jaritz}, \& {Biernat}}]{erkaev2007roche-lobe}
{Erkaev}, N.~V., {Kulikov}, Y.~N., {Lammer}, H., {et~al.} 2007, \aap, 472, 329,
  \dodoi{10.1051/0004-6361:20066929}

\bibitem[{{Erkaev} {et~al.}(2016){Erkaev}, {Lammer}, {Odert}, {Kislyakova},
  {Johnstone}, {G{\"u}del}, \& {Khodachenko}}]{erkaev2016MNRAS.460.1300E}
{Erkaev}, N.~V., {Lammer}, H., {Odert}, P., {et~al.} 2016, \mnras, 460, 1300,
  \dodoi{10.1093/mnras/stw935}

\bibitem[{{Fossati} {et~al.}(2017){Fossati}, {Erkaev}, {Lammer}, {Cubillos},
  {Odert}, {Juvan}, {Kislyakova}, {Lendl}, {Kubyshkina}, \&
  {Bauer}}]{fossati2017aeronomical_constr}
{Fossati}, L., {Erkaev}, N.~V., {Lammer}, H., {et~al.} 2017, \aap, 598, A90,
  \dodoi{10.1051/0004-6361/201629716}

\bibitem[{{Gupta} \& {Schlichting}(2020)}]{gupta_schlichting2020core-powered}
{Gupta}, A., \& {Schlichting}, H.~E. 2020, \mnras, 493, 792,
  \dodoi{10.1093/mnras/staa315}

\bibitem[{{Kubyshkina} {et~al.}(2018{\natexlab{a}}){Kubyshkina}, {Fossati},
  {Erkaev}, {Johnstone}, {Cubillos}, {Kislyakova}, {Lammer}, {Lendl}, \&
  {Odert}}]{kubyshkina2018grid}
{Kubyshkina}, D., {Fossati}, L., {Erkaev}, N.~V., {et~al.} 2018{\natexlab{a}},
  \aap, 619, A151, \dodoi{10.1051/0004-6361/201833737}

\bibitem[{{Kubyshkina} {et~al.}(2018{\natexlab{b}}){Kubyshkina}, {Fossati},
  {Erkaev}, {Cubillos}, {Johnstone}, {Kislyakova}, {Lammer}, {Lendl}, \&
  {Odert}}]{kubyshkina2018approx}
---. 2018{\natexlab{b}}, \apjl, 866, L18, \dodoi{10.3847/2041-8213/aae586}

\bibitem[{{Watson} {et~al.}(1981){Watson}, {Donahue}, \&
  {Walker}}]{watson1981energy-limited}
{Watson}, A.~J., {Donahue}, T.~M., \& {Walker}, J.~C.~G. 1981, \icarus, 48,
  150, \dodoi{10.1016/0019-1035(81)90101-9}

\bibitem[{{Yi} {et~al.}(2001){Yi}, {Demarque}, {Kim}, {Lee}, {Ree}, {Lejeune},
  \& {Barnes}}]{yi2001stellar-models}
{Yi}, S., {Demarque}, P., {Kim}, Y.-C., {et~al.} 2001, \apjs, 136, 417,
  \dodoi{10.1086/321795}

\end{thebibliography}
\bibliographystyle{aasjournal}



\end{document}